\documentstyle[11pt,cs11,html,epsf]{article}

\begin{document}

\title{Chromospheric Activity in Metal-Poor Dwarfs}

\author{R. C. Peterson}
\affil{Astrophysical Advances, Palo Alto, CA  94301; UCO/Lick Obs., Santa Cruz, CA  95064}

\author{C. J. Schrijver}
\affil{Lockheed Martin, Palo Alto, CA  94304-1191}



\begin{abstract}
We have obtained echelle spectra with the Hubble Space Telescope (HST) of the MgII 2800\,\AA\ region of ten stars 
whose metallicities range from 1/300 to 1/3 that of the Sun, and whose space velocities suggest a halo or old thick-disk 
origin.  Spectra of all ten show double-peaked emission in the MgII core, very much like the quiet Sun. 
A half-dozen apparently non-rotating stars were observed more than once, and show at most a low level of variability in the 
emission flux, comparable to that of quiet stars of solar metallicity.
For four stars, we have obtained Lyman\,$\alpha$ spectra at 0.2\,\AA\ resolution; all four show emission.
The data thus strongly suggest that chromospheric activity at a minimum level is present in all stars 
of near-solar temperature, regardless of age or metallicity. While this points to non-magnetic sources such as 
acoustic waves, a contribution from globally-organized magnetic fields is possible.   
A longer series of MgII and Lyman\,$\alpha$ observations is needed to constrain this.
\end{abstract}


%


\medskip
\noindent {\bf Background: quiet chromospheres in old stars.} Whether chromospheric activity in quiet stars of solar metallicity and temperature is due to a globally-organized magnetic field generated by internal stellar rotation remains an open question. Among active stars, such a ``dynamo'' is implicated as the dominant source of a chromosphere by two separate observations: by the fact that the brightest chromospheric emission on the surface of the Sun appears in regions near sunspots (Lemaire \& Skumanich 1973; Schrijver et al.\ 1989; Brekke et al.\ 1991), and by the positive correlation  between emission and surface rotation found among cool stars generally (Noyes et al.\ 1984). However, both observation and theory suggest that weak chromospheric emission may still produced in the absence of a dynamo. If generated by convection, then chromospheric emission at some low, roughly constant level should appear in all stars with convection zones. We test these ideas here by examining chromospheric MgII emission over a three-year period for a half-dozen apparently non-rotating solar-type stars believed to be older than the Sun, for which the rotation generating a magnetic dynamo has had the longest chance to spin down. As a more sensitive diagnostic of dynamo-related activity, we obtained a single Lyman\,$\alpha$ spectrum for four stars.

On the Sun, most chromospheric emission is confined to regions of strong magnetic fields, in sunspot-related plages and in the magnetic network covering the entire surface. Most of that field changes in strength with the solar magnetic cycle, although there is almost always an area of network near the equator that shows a low, cycle-independent magnetic flux density. Whatever drives the network field, however, 
the emission produced in even its weakest regions is stronger than the emission generated in the interior regions of the cells it defines, where the magnetic field is weak and turbulently advected. Only in these cell interiors does solar chromospheric emission reach the minimum or ``basal'' level found among cool stars generally (Schrijver 1987; Rutten et al.\ 1991). Theoretically, as reviewed by Schrijver (1995) and Narain \& Ulmschneider (1996), this basal emission may be produced by acoustic waves generated by convection. Acoustic waves contain enough energy (Ulmschneider 1990), and their inhomogeneity in space and time leads naturally to the small-scale spatial and temporal variations seen on the solar surface (Carlsson \& Stein 1994). 

Whether normal metal-poor dwarfs always show such emission is debatable, since it is rarely detected in the CaII line normally observed. B\"ohm-Vitense (1982), Dupree, Hartmann, \& Smith (1990), Judge \& Stencel (1991), and Dravins et al.\ (1993) found CaII emission in old stars, but these are subgiants and giants. Among metal-poor dwarfs, one star with CaII emission proved to be a tidally-locked binary (Peterson et al.\ 1980). Two main-sequence stars were found by Duncan et al.\ (1991) to have cyclical variations in CaII, suggestive of a dynamo. However, the hotter one, HD 106516, shows finite rotation, and the cooler one, HD 103095 (Groombridge 1830), is much cooler than the Sun (Peterson \& Carney 1979).  Recently, Smith \& Churchill (1998) have found CaII emission in a number of metal-poor stars, again mostly cooler than the Sun.


\medskip \noindent {\bf MgII fluxes and line profiles.} The CaII emission of the oldest stars may be weakened by low metallicity to the point of falling below detection limits. The MgII emission features should be stronger: magnesium is about fifteen times more abundant than calcium, and the stellar continuum background near MgII, at 2800\,\AA, is substantially lower than that near CaII at 3933\,\AA.

Consequently, we began to obtain MgII spectra to explore systematically the chromospheres of old solar-temperature dwarfs. We used the Hubble Space Telescope (HST) with the Goddard High Resolution Spectrograph (GHRS). During Cycle 5 in 1995--96 we obtained spectra in snapshot mode (with 3 -- 7\,min exposure times) of a 50\,\AA\ region centered on 2800\,\AA\ at a FWHM resolution of 0.01\,\AA\ (4.4 pixels). The ten targets are the brightest stars with temperatures 5700K -- 6500K for which metallicity [Fe/H] $< -0.5$, i.e.\ less than one-third solar, and space velocity with respect to the local standard of rest Z$_{\rm LSR}$ $> 65$ km/s. They are HD 19445, HD 84937, HD 94028, HD 106516, HD 114762, HD 140283, HD 184499, HD 194598, HD 201891, and HD 219617. The list includes extreme halo stars as well as mildly metal-poor members of the thick disk. Most were observed by Hipparcos; several proved to be subgiants. According to Carney et al.\ (1994), seven have radial velocities V$_R$ constant to 1 km/s, and three are binaries, as described in the discussion at the end. One of the binaries, HD 106516, is rotating; we defer its discussion to a separate paper, considering here only the nine stars which show no rotation ($v$ sin $i$ $<$ 6 km/s), whose chromospheres are expected to be very quiet.

As reported by  Peterson \& Schrijver (1997), all of the nine stars show spectra with weak to moderate MgII emission. Profiles are seen to be double-peaked, except where the ever-present absorption due to the local interstellar medium (LISM) overlaps in velocity with the stellar feature. As in the Sun, the blue peak is $\sim$20\% stronger than the red, and the $k$ line at 2795.5\,\AA\ is $\sim$20\% stronger than the $h$ line at 2802.7\,\AA. Peak separations are $\le$ 0.5\,\AA, rather narrower than expected from the Wilson-Bappu relation for MgII. Figure 1 illustrates these features by plotting all the MgII spectra obtained for two dwarfs at the extremes of the metallicity range: HD 114762 and HD 19445, with metallicity 1/6 and $<$ 1/100 solar respectively. LISM absorption is expected to occur at wavelengths displaced blueward by 0.55\,\AA\ for HD 114762 and redward by 1.49\,\AA\ for HD 19445, and such features are indeed present. The spectra are shifted by the stellar radial velocities to rest velocity and plotted on an air wavelength scale.

Emission fluxes are somewhat metal-dependent. Fluxes of mildly metal-poor stars are comparable to that of HD 128620 ($\alpha$ Cen A), a well-studied, chromospherically quiet (Ayres et al.\ 1995), somewhat metal-rich dwarf of solar type and age (Furenlid \& Meylan 1990), while those of the most metal-poor stars are even lower, near the basal flux level of Rutten et al. Since 
the majority of stars show emission at higher levels than basal, these data alone do not rule out the presence or even dominance of a contribution from global magnetic fields.

In hopes of detecting variations in MgII emission in stars with strong activity cycles, we reobserved MgII three years later. HST's Space Telescope Imaging Spectrograph (STIS) echelle in snapshot mode (with 3.3 -- 10\,min exposures) produced spectra of comparable resolution and signal-to-noise. Six stars were observed in 1998--1999 during Cycle 7: the visual binary HD 219617 was dropped, and HD 140283 and HD 194598 never made it into the snapshot schedule. HD 19445, HD 84937, and HD 114762 were observed twice more.

As before, all spectra show double-peaked MgII emission. Comparison of emission fluxes in GHRS and STIS spectra proved straightforward, in that all spectra show flux levels of absorption wings of the MgII absorption line profiles that are constant to 1\% between GHRS and STIS. This was undoubtedly facilitated by the fact that both sets of observations used relatively large square apertures ($2^{\prime\prime}$ and $0.2^{\prime\prime}$ respectively) to minimize setup time. 

No statistically significant variation in MgII flux is seen in Figure 1 or elsewhere. With few exceptions (such as the first and last exposures of HD 114762), the level of emission flux is constant to 15\% for the height of an individual subpeak and half this size for the average of $h$ and $k$ line emission combined. Thus these data do not rule out an intrinsic variation at the level of 10 -- 15\%, the size seen in $\alpha$ Cen A and in solar-metallicity F-G dwarfs in general (see Figure 12$a$ of Ayres et al.). Changes of this size could be dynamo-related or not. They are routinely seen in levels of ``quiet'' activity on the solar surface. Moreover, especially in metal-poor stars but even in $\alpha$ Cen A, the contrast of the emission above the absorption line background is low, and a change in this background could cause a change of this size in overall central flux.  

\begin{figure}
\plotone{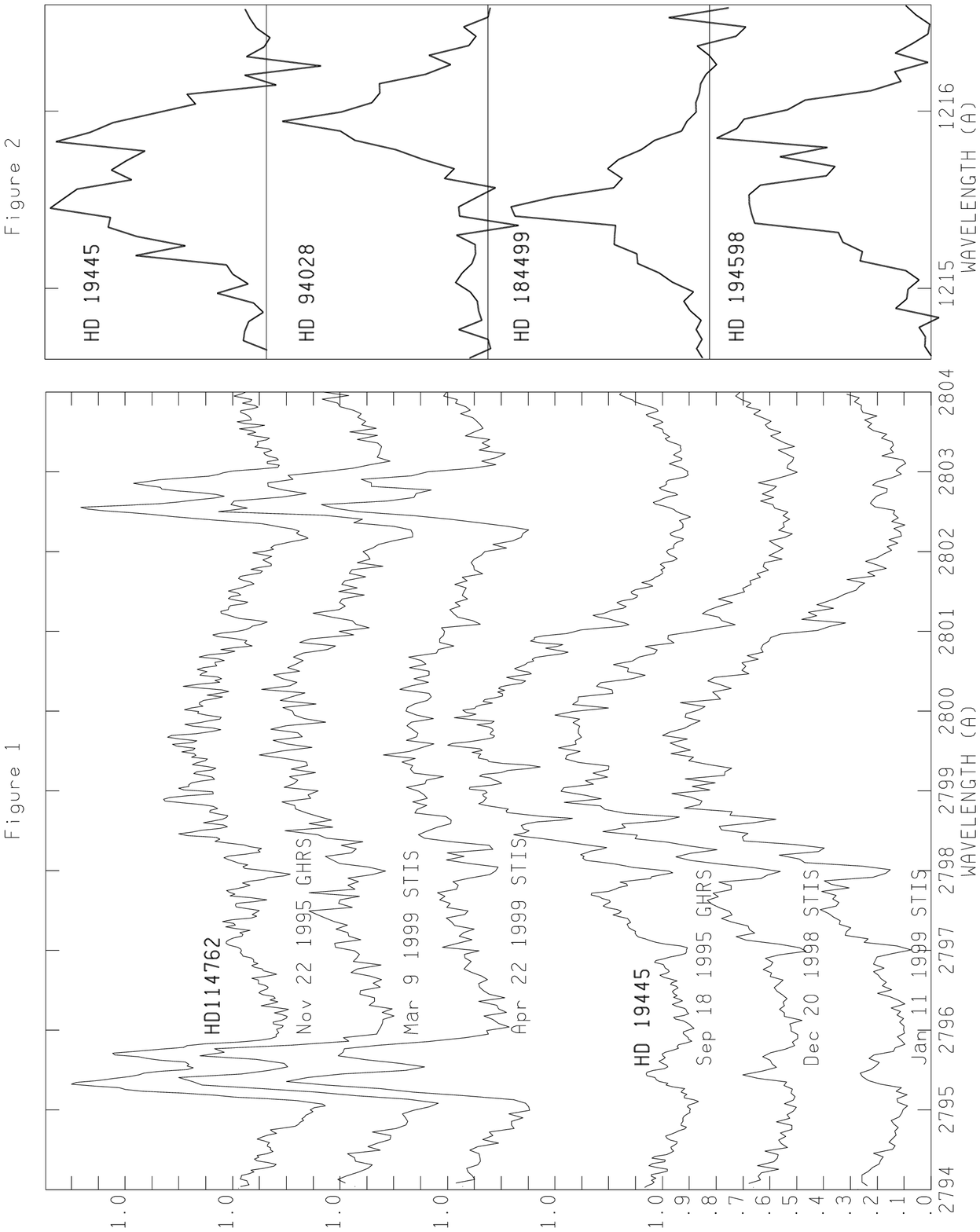} 
\end{figure}
 

\medskip \noindent {\bf A new approach: Lyman\,$\alpha$ line profiles and fluxes.} We have also acquired observations of Lyman\,$\alpha$ to look for asymmetries in the profile or other indicators of enhanced and/or dynamo-generated activity. The Sun is again a useful guide. Spectra depicted by Brekke et al.\ (1991) show that in the quiet Sun, the Lyman\,$\alpha$ profile morphologically resembles that of MgII: it is double-peaked, but with near equality of the blue and red peaks and with a peak separation comparable when measured in \AA\ and not in km/s. In passing from quiet regions to active regions, light bridges, and sunspots, the solar Lyman\,$\alpha$ profile increases in both intensity and breadth, and becomes progressively more irregular in shape. Thus the Lyman\,$\alpha$ emission flux and profile shape in metal-poor stars might reveal more directly whether sunspot-associated activity is present on their surfaces. Even a small fraction of sunspot-related activity might be detected, since Lyman\,$\alpha$ is vastly more sensitive to emitting material than MgII. Hydrogen is 30,000 times more abundant than magnesium in the Sun (more so in metal-poor stars) and there is no absorption spectrum at 1216\,\AA\ in stars of 6000K (e.g.\ Ayres et al.). 

Unfortunately, Lyman\,$\alpha$ is also an incredibly sensitive tracer of the LISM. In their study of the interstellar deuterium abundance in $\alpha$ Cen A, the closest stellar system beyond the Sun, Linsky \& Wood (1996) show a Lyman\,$\alpha$ profile in which the LISM has absorbed the central 0.5\,\AA, and their reconstituted profile suggests a loss of 2/3 of the flux to the LISM. Similarly, Judge \& Stencel (1991) remark that the LISM renders the Lyman\,$\alpha$ fluxes of the relatively nearby giants Arcturus and $\alpha$ Tau uncertain by 50\%. With the seriousness of this effect in mind, Landsman \& Simon (1993) presented a model for correcting the observed flux for interstellar attenuation with their catalog of stellar IUE Lyman\,$\alpha$ fluxes. 

Our approach is to bypass the LISM altogether by observing stars of high radial velocity. We selected six stars, four with negative radial velocities that shift their spectra $> 150$ km/s from the LISM, and the two with the largest positive velocity shifts for comparison. Their generally low metallicities should not matter here since hydrogen is the emitter. The stars are faint, but their velocities minimize the contribution from geocoronal Lyman\,$\alpha$, and lower resolution can be used. We chose the STIS G140M first-order grating, whose 0.2\,\AA\ 
4-pixel resolution should easily resolve the separate peaks, whose long-slit operation allows geocoronal subtraction, and whose throughput at Lyman\,$\alpha$ is 2.5\,$\times$ that of the echelles. During Cycle 7 four snapshot spectra were taken. The stars (with heliocentric radial velocities and positions in Galactic coordinates $\ell$, $b$) are HD 19445 ($-140$ km/s, 157\deg, $-$27\deg), HD 94028 (+65 km/s, 220\deg, +61\deg), HD 184499 ($-167$ km/s, 67\deg, +6\deg), and HD 194598 ($-248$ km/s, 53\deg, $-$16\deg). 

Figure 2 shows the resulting spectra shifted to rest velocity and plotted on a vacuum wavelength scale. All four show Lyman\,$\alpha$ emission. The spectra of HD 19445 and HD 194598 are double-peaked, as expected from their high velocities. The blue peak appears to be suppressed by the LISM in the spectrum of HD 94028, consistent with its moderate positive velocity. Unexpectedly, only a blue peak is present in the HD 184499 spectrum, despite its high velocity. Whether this is due to a high-velocity ISM cloud or to an intrinsic variation in the stellar profile cannot be judged at present, given the small sample of spectra and their relatively low signal-to-noise. Note that HD 184499 is a subgiant with $M_v$ = 4.1, according to its Hipparcos parallax (Reid 1998), so is about 30pc away.


\medskip \noindent {\bf Conclusions and future work.} Our work has shown that quiet, metal-poor stars of near-solar temperature do all have chromospheres.  MgII or Lyman\,$\alpha$ emission is detected in every spectrum. 

The mechanism of production of chromospheric emission in old, metal-poor stars is apparently similar to that operating in quiet regions of the Sun. All the stellar MgII profiles are double-peaked, like those of the quiet Sun but with somewhat smaller peak separations, especially in extremely metal-poor stars. The level of MgII emission is quite a bit weaker in these stars as well, approaching the empirical basal flux, but is only mildly weaker than solar in the not-so-metal-poor stars. As seen in Figure 1, however, the range in emission flux is much less than the factor of 100 in metallicity. Whatever the mechanism responsible for chromospheric line-core reversals, opacity effects on flux levels are largely suppressed. This has been both predicted (Cuntz, Rammacher, \& Ulmschneider 1994) and observed (Dupree et al.\ 1990; Dupree \& Smith 1995) for giant stars, and is demonstrated here for solar-type dwarfs and subgiants also.

There is no evidence for global-dynamo-generated chromospheric activity on any of these nonrotating stars, but the presence of a dynamo cannot be ruled out. MgII variability is low, generally $<$ 10\%, within the uncertainties of measurement and consistent with the low variability seen in quiet solar-neighborhood stars such as $\alpha$ Cen A. This confirms that these are indeed quiet stars. However, there is still room for those who believe that global magnetic fields could be the dominant mechanism of chromospheric emission. The largest changes might have been missed by the three-year spacing of observations, given that the solar cycle is 11 years and that the period and phase of variability in these stars is unknown. Thus a third set of HST observations in both MgII and Lyman\,$\alpha$ will be proposed for cycle 10, another three or four years later. Where possible, stars of a wider range of temperatures and metallicities will be included, to assess the run of line width with temperature and metallicity as well as with luminosity.

Lyman\,$\alpha$ emission is seen in all cases, but its profile is not yet well defined from this limited dataset.
Two of the Lyman\,$\alpha$ profiles are double-peaked; the high stellar radial velocities have shifted the emission out from under LISM absorption. Single peaks are found in two cases, one a high-velocity star but one a star of low velocity. For the latter profile to be affected by LISM absorption, an ISM cloud is required with velocity near 100 km/s within 30pc, an unprecedented circumstance (Wakker \& van Woerden 1997). Should this be case nonetheless, it is reasonable to assume that all Lyman\,$\alpha$ profiles of solar-temperature dwarfs are intrinsically double-peaked with similar peak separation. Demonstrating this from additional Lyman\,$\alpha$ spectra of high-velocity stars is a high priority for future work. It has the potential to provide constraints on chromospheric formation, and to refine future measurements of interstellar deuterium abundances by possibly providing a standard template for the intrinsic Lyman\,$\alpha$ emission profile of a slowly rotating, otherwise inactive F-G dwarf or subgiant.

With the assumption that all four Lyman\,$\alpha$ profiles show two peaks of similar height and spacing, total apparent Lyman\,$\alpha$ fluxes can be calculated. The distance-independent Lyman\,$\alpha$-to-MgII flux ratio is found to depend strongly on stellar abundance, being smallest at 0.27 for HD 184499 at [Fe/H] = $-0.7$ (after LISM correction) and largest at 1.22 for HD 19445 at [Fe/H] = $-2.1$. The scatter is consistent with observational error alone.  This variation in the Lyman\,$\alpha$-to-MgII flux ratio suggests that the cooling of the lower chromosphere is reduced in metal-poor stars, which may affect their structure. Moreover, especially in the most metal-poor stars, the Lyman\,$\alpha$ emission flux by itself evidently provides a substantial contribution to the total energy lost by a stellar envelope to chromospheric emission. Additional Lyman\,$\alpha$ observations of high-velocity stars could provide a meaningful lower bound to the energy input required to produce a chromosphere, an important constraint for chromospheric models. 

Using Hipparcos distances and stellar gravities derived by assuming a mass 0.8 that of the Sun, intrinsic Lyman\,$\alpha$ fluxes may be derived. The resulting values show no [Fe/H] dependence, and are more constant from star to star than the MgII fluxes. The star with the highest Lyman\,$\alpha$ flux is the subgiant HD 184499, hinting that surface Lyman\,$\alpha$ flux may be the most constant quantity. This remains to be established from future Lyman\,$\alpha$ observations.


%

%
%
%
%

\acknowledgments

We thank B. Carney, H. Lanning, M. McGrath, K. Peterson, and T. Royle for invaluable aid 
in planning these observations, and C. Zwaan, R. Rutten, G. Smith, and R. Kraft for
helpful conversations. This research has made use of the SIMBAD 
database, operated at CDS, Strasbourg, France. Support for this work was provided by NASA 
through grant numbers GO-05869.01-94A, GO-07395.01-96A, and GO-07402.01-96A from the Space 
Telescope Science Institute, which is operated by the Association of Universities for Research 
in Astronomy, Inc., under NASA contract NAS5-26555.

\begin{question}{M. Cuntz}
I just want to point out that P. Ulmschneider and his group are in the process
of revisiting the ``chromospheric basal flux line'' considering different
metal abundances. This will allow us to judge whether there is still
room for magnetic heating in stars with very slow rotation.
\end{question}

\begin{question}{K. Strassmeier}
Are there binaries in this sample?
\end{question}
\begin{answer}{R. C. Peterson}
We know of three binaries among our targets. 
HD 114762 is a velocity variable of period 83.9 days with an unseen companion, 
either a brown-dwarf or a planet (Latham et al.\ 1989). 
HD 219617 (ADS 16644A,B) is a visual binary with very similar components separated by $0.7$\arcsec.
HD 106516 is a velocity variable with a period of 845 days 
in an orbit of virtually zero eccentricity (Latham et al.\ 1992). 
It has low but detectable rotation: $v$ sin $i \approx 8$ km/s, according to SIMBAD. 
These properties, and the fact that it is situated slightly above the metal-poor main-sequence turnoff, suggest 
it has undergone mass exchange in a blue straggler system. We included it as a rare case of a possibly active 
old, metal-poor star, and discuss its spectra in a separate paper. 
\end{answer}

\begin{question}{C. Allende Prieto}
Wouldn't it be useful to search for the magnesium emission in IUE spectra?
\end{question}
\begin{answer}{R. C. Peterson}
Four of our dwarf targets and several other metal-poor stars 
have been observed at high resolution with the International Ultraviolet Explorer (IUE) in the MgII region.
Unfortunately, because of limited dynamic range and difficult background subtraction,
the signal-to-noise was not sufficient to detect the very weak emission of 
our targets HD 19445, HD 84937, and HD 140283, all with $V > 7$. 
Even for the $V$ = 6 giant HD 122563, emission is only marginally detected.
All three of the remaining metal-poor stars known with $V < 7$ 
do indeed show double-peaked MgII emission. 
Two of these stars are HD 106516 and HD 103095, 
discussed above and in the text. The third is the field blue horizontal branch star HD 161817. 
It illustrates that even a late A star can exhibit chromospheric activity, presumably because 
it is cool enough for strong convection to occur (Castelli, Gratton, \& Kurucz 1990).
\end{answer}

\end{document}